\documentclass[3p,times,twocolumn]{elsarticle}

\usepackage{ecrc}

\volume{00}

\firstpage{1}

\journalname{Nuclear Physics B Proceedings Supplement}

\runauth{}

\jid{nuphbp}

\jnltitlelogo{Nuclear Physics B Proceedings Supplement}

\usepackage{amssymb}

\usepackage{units}
\usepackage{amsmath}

\usepackage[figuresright]{rotating}

\begin{document}

\begin{frontmatter}



\dochead{}

\title{An Improved Measurement of Electron Antineutrino Disappearance at Daya Bay}


\author{David M. Webber, on behalf of the Daya Bay Collaboration}

\address{University of Wisconsin, Madison}

\begin{abstract}
The theory of neutrino oscillations explains changes in neutrino flavor, count rates, and spectra from solar, atmospheric, accelerator, and reactor neutrinos. These oscillations are characterized by three mixing angles and two mass-squared differences. The solar mixing angle, $\theta_{12}$, and the atmospheric mixing angle, $\theta_{23}$, have been well measured, but until recently the neutrino mixing angle $\theta_{13}$ was not well known. The Daya Bay experiment, located northeast of Hong Kong at the Guangdong Nuclear Power Complex in China, has made a precise measurement of electron antineutrino disappearance using six functionally-identical gadolinium-doped liquid scintillator-based detectors at three sites with distances between 364 and 1900 meters from six reactor cores. This proceeding describes the Daya Bay updated result, using 127 days of good run time collected between December 24, 2011 and May 11, 2012.  For the far site, the ratio of the observed number of events to the expected number of events assuming no neutrino oscillation is $0.944 \pm 0.007(\mathrm{stat}) \pm 0.003(\mathrm{syst}).$  A fit for $\theta_{13}$ in the three-neutrino framework yields $\sin^2 2\theta_{13} = 0.089 \pm 0.010(\mathrm{stat}) \pm 0.005(\mathrm{syst}).$
\end{abstract}


\end{frontmatter}


\section{Introduction}
In the Standard Model of particle physics, the three neutrino flavors $\nu_e,$ $\nu_\mu,$ $\nu_\tau,$ corresponding to the three generations of matter, are massless.  However, the discovery of neutrino flavor oscillation \cite{SuperK_1998, SNO_2002} has shown that neutrinos must have mass. 
Neutrinos propagate in definite mass eigenstates, denoted $\nu_1,$ $\nu_2,$ $\nu_3,$ and the mixing between the mass and flavor eigenstates is described by three mixing angles and one complex phase.  Until this year, the smallest mixing angle, $\theta_{13}$, was not known.  Previous experiments observing reactor antineutrinos showed that $\sin^2 2\theta_{13}<0.17$ at 90\% confidence \cite{Chooz1, Chooz2}.  More recently, beam neutrino \cite{MINOS, T2K} and reactor antineutrino \cite{DoubleChooz} experiments showed hints that $\sin^2 2\theta_{13}$ is nonzero with significances between 1 and 2.5~$\sigma$.
In March 2012, the Daya Bay experiment released results showing that $\sin^2 2\theta_{13}$ is nonzero to better than $5~\sigma$ \cite{DayaBay_PRL}, and this result was confirmed by the RENO experiment \cite{RENO_2012}.

After this announcement, the Daya Bay experiment has updated its result including an additional 77 days ($2.5\times$) exposure \cite{DayaBay_improved}.  This proceeding describes an overview of neutrino oscillation and the Daya Bay experiment, and presents the updated result.

\section{Neutrino Oscillation}
As an introduction to neutrino oscillations, consider the simple case of two neutrinos, with flavor eigenstates $\nu_a$ and $\nu_b$ and mass eigenstates $\nu_1$ and $\nu_2$, related by one mixing angle $\theta$.
\begin{equation}
\left (
\begin{array}{c}
\nu_a \\
\nu_b
\end{array}
\right )
=
\left (
\begin{array}{rr}
\cos \theta  & \sin \theta \\
-\sin \theta & \cos \theta
\end{array}
\right )
\left (
\begin{array}{c}
\nu_1 \\
\nu_2
\end{array}
\right )
\end{equation}
Neutrinos initially in a pure $\nu_a$ state will oscillate in and out of a pure $\nu_a$ state following the wave function
\begin{equation}
\Psi_{\nu_a}(x,t) = f(x,t) \sum_i U_{ai}e^{-i(m_i t/2E)}.
\end{equation}
The ``survival probability'' for a neutrino initially in a pure $\nu_a$ state to remain in the $\nu_a$ state is
\begin{equation}
P(\nu_a\rightarrow\nu_a) = 1 - \sin^2(2\theta) \sin^2 \left( 1.27 \Delta m_{21}^2 \frac{L}{E_\nu} \right),
\end{equation}
where $\theta$ governs the amplitude of the oscillation, $\Delta m_{21}^2 (eV^2) = m_2^2 - m_1^2 $
governs the frequency of oscillation, $L$ (km) is the propagation distance, and $E_\nu$ (GeV) is the neutrino energy.

By extending this framework to three neutrinos \cite{MNSmatrix} and making the approximation $\Delta m^2_{32} \approx \Delta m^2_{31} \approx \Delta m^2_{\mathrm atm},$
one can write the survival probability of electron antineutrinos as

\begin{equation}
\label{eqn:psur}
\begin{split}
P_{\mathrm{sur}} \approx 1 &- \sin^2 2\theta_{13} \sin^2 \left( \Delta m^2_{32} \frac{L}{4E} \right) \\
 &- \sin^2 2\theta_{12} \cos^4 2\theta_{13} \sin^2\left(\Delta m^2_{21} \frac{L}{4E}\right).
\end{split}
\end{equation}

From equation~\ref{eqn:psur}, it is clear that two terms dominate the survival probability. 
For a fixed antineutrino energy range, the first term contains the larger $\Delta m^2_{32}$ and becomes dominant at smaller baselines.  The second term depends on the smaller $\Delta m^2_{21}$ and applies to the measurement of $\theta_{12}$ at larger baselines.  The latter term has been measured by KamLAND \cite{Kamland_2008}, and the former is well suited to a $\approx${2}~{km} baseline reactor antineutrino experiment.

\section{The Daya Bay Experiment}
\begin{figure}[h]
\begin{center}
\includegraphics[width=0.95\columnwidth]{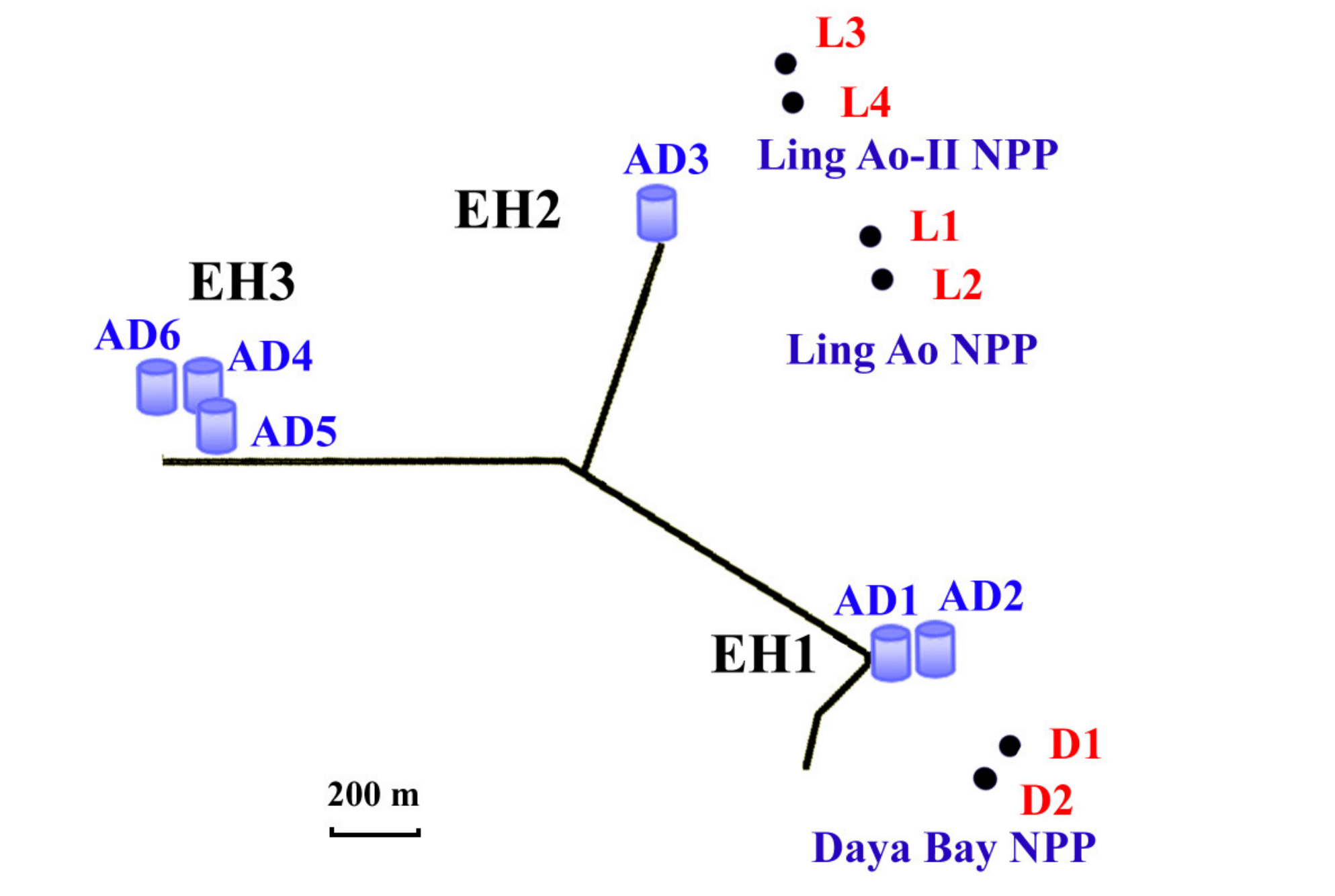}
\caption{The layout of the Daya Bay experiment.  The six reactors (D1,D2,L1,L2,L3,L4) are located at the Daya Bay (D.B.), Ling Ao (L.A.), and Ling Ao-II (L.A.II) nuclear power plants.  The Daya Bay antineutrino detectors (ADs) are located underground in three experimental halls (EHs). Underground tunnels connect the three sites and provide access from the surface.\label{fig:sitelayout}}
\end{center}
\end{figure}

The Daya Bay experiment takes place at the Daya Bay nuclear power complex near Shenzhen, China.
There are 6 reactors spaced in pairs along the coast in three nuclear power plants.  Each reactor is capable of producing 2.9~GW thermal power, for a total site power of 17.4~GW.  
The reactor companies provide time-averaged information about the reactor power and the isotope fractions to the experiment; uncertainties on these quantities are given in Table~\ref{tbl:reactor_uncertainties}.
\begin{table}
\begin{center}
\caption{Reactor Uncertainties\label{tbl:reactor_uncertainties}.  Note that
for a relative near-far measurement, only the uncorrelated uncertainties contribute.}
\begin{tabular}{l l l l}
\hline
\hline
\multicolumn{2}{c}{Correlated} & \multicolumn{2}{c}{Uncorrelated} \\
\hline
Energy/fission & 0.2\% & Power & 0.5\% \\
IBDs$^\dagger$/fission    & 3\%   & Fission fraction & 0.6\% \\
               &       & Spend fuel & 0.3\% \\
\hline
Combined & 3\% & Combined & 0.8\%  \\
\hline
\hline
$^\dagger$inverse beta decays
\end{tabular}
\end{center}
\end{table}
Overall, the nuclear power complex produces $3\times10^{21}$ antineutrinos per second. During the collection of the December 24 -- May 11 dataset, 
there were 6 detectors in installed in three experimental halls, as shown in Figure~\ref{fig:sitelayout}.  The detectors are installed underground with baselines and overburdens summarized in Table~\ref{tbl:baselines}.

\begin{table}[h]
\begin{center}
\caption{Experimental hall overburden (m.w.e.) and approximate baselines to reactor cores (m). 
An overview of the experimental site layout is shown in Figure \ref{fig:sitelayout}.
\label{tbl:baselines}}
\begin{tabular}{l c c c c}
\hline
\hline
 & Overburden & D.B. & L.A. & L.A.II \\
\hline
EH1 & 280 & 360  & 860 & 1310 \\
EH2 & 300 & 1250 & 480 & 530 \\
EH3 & 880 & 1910 & 1540 & 1550 \\
\hline
\hline
\end{tabular}
\end{center}
\end{table}



The Daya Bay antineutrino detectors (ADs) have 3 cylindrical nested fluid volumes separated by two acrylic vessels.  The innermost volume 
holds 20~tons of linear-alkyl-benzene-based liquid scintillator, doped with 0.1\% gadolinium.  This gadolinium-doped liquid scintillator (GdLS) optimizes the detection of the 
 the inverse beta decay (IBD) process $\bar{\nu}_e + p \rightarrow n + e^+$, where the positron carries most of the neutrino energy and is tagged by neutron capture on gadolinium. 
The intermediate volume contains undoped liquid scintillator (LS) and serves as a gamma-ray catcher for light escaping from the central volume.  The outer volume contains a non-scintillating mineral oil (MO) buffer. Scintillation light is collected by reflectors on the top and bottom of the detector and by 192 photomultiplier tubes on the walls of the AD.

\begin{figure}[h]
\begin{center}
\includegraphics[width=0.95\columnwidth]{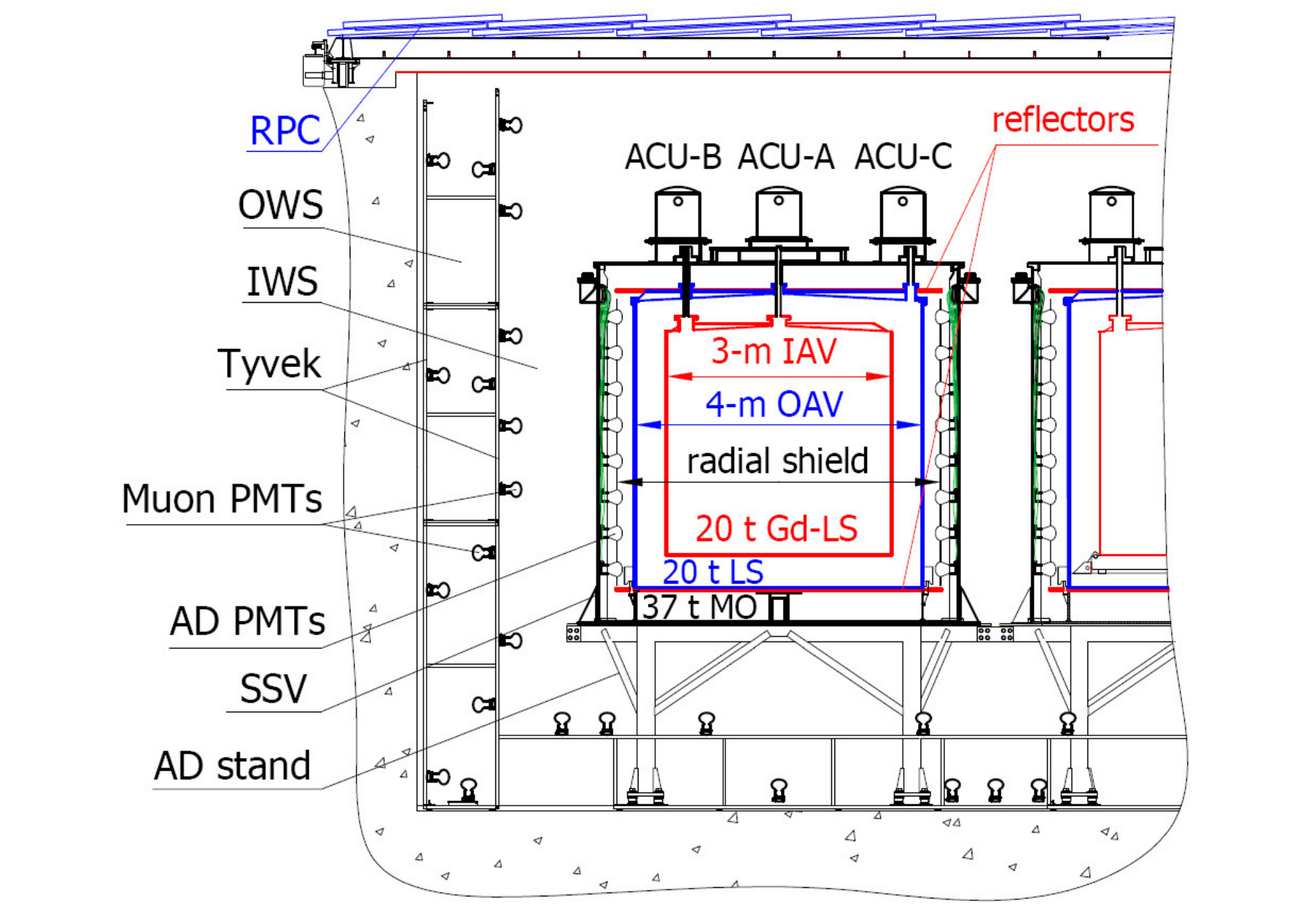}
\caption{Schematic of installed antineutrino detector (AD). The AD has three cylindrical nested fluid volumes, filled with gadolinium-doped liquid scintillator, undoped liquid scintillator, and mineral oil.  Top and bottom reflectors guide scintillation light into 192 8-inch photomultiplier tubes (PMTs) on the walls of the detector.  The ADs are installed on supports in a water pool instrumented with PMTs.  The water pool and the RPCs serve as a cosmic ray veto. Automated calibration units (ACUs) periodically lower sources into the detector for calibration.\label{fig:AD}}
\end{center}
\end{figure}

The detectors are constructed on the surface in a clean environment, and then transported underground for filling.  The detector is filled simultaneously from three fluid reservoirs to keep the fluid levels equal and to minimize the stress on the acrylic vessels.  After filling, the detectors are transported to one of the three experimental halls, where they are placed in a water pool.  Each water pool is filled with ultra-pure (15~megohm-cm) water and serves the dual function of radiation shield and active muon veto.  The water pool is covered with a light-tight cover and an array of resistive-plate chambers (RPCs) \cite{DYB_RPC} is rolled over the top as an additional cosmic ray veto.  A schematic of an AD installed in the water pool is shown in Figure~\ref{fig:AD}.
The detectors are functionally identical and have excellent consistency, as studied in a side-by-side comparison of the first two detectors in hall 1 \cite{DYB_AD12}.

\section{Event Selection}
Although care is taken to keep radioactive materials out of the detectors, antineutrino interactions in the detector are rare compared to ambient radioactive backgrounds.  With a trigger threshold of 0.4~MeV, the single-event rate is $\approx${250}~{Hz} ($\approx${140}~{Hz}), whereas the neutrino interaction rate is $\approx${650}{/day} ($\approx${75}{/day}) at the near (far) sites.
The Daya Bay analysis uses coincident events to reduce this background.  Specifically, the detectors are optimized for observing the prompt positron and delayed neutron of the IBD process.  The positron carries most of the energy of the neutrino, and it thermalizes and annihilates quickly.  
The average neutron capture time in GdLS is {28}~{$\mu$s}.  This neutron capture 
provides an easily identified delayed signal which tags the prompt events.  After rejecting flashing PMT events and applying a muon veto, IBD candidates are selected from all the prompt-delayed coincidence events which have;
\begin{itemize}
\item prompt energy between 0.7 and 12 MeV,
\item delayed energy between 6 and 12 MeV,
\item time between prompt and delayed events between 1 and 200~$\mu s$,
\item no other signals {400}~{$\mu s$} before or {200}~{$\mu s$} after the delayed neutron event.
\end{itemize}
The efficiency and uncertainty of these cuts, as well as other detector-related uncertainties, is summarized in Table~\ref{tbl:detector_uncertainties}.

\begin{table}[ht]
\tabcolsep=0.11cm
\begin{center}
\caption{Detector Uncertainties\label{tbl:detector_uncertainties}.  The spill-in efficiency is larger than 100\% since more IBD neutrons drift inward into the central GdLS target volume than outward into the LS volume. 
For a relative near-far measurement, only the uncorrelated uncertainties contribute.}
\small
\begin{tabular}{l l l l}
\hline
\hline
                      & Efficiency & Correlated & Uncorrelated \\
\hline 
Target Protons        & --         &  0.47\%    & 0.03\%       \\
Flasher Cut           & 99.98\%    &  0.01\%    & 0.01\%       \\ 
Delayed Energy Cut    & 90.9\%     &  0.6\%     & 0.12\%       \\
Prompt Energy Cut     & 99.88\%    &  0.10\%    & 0.01\%       \\
Multiplicity Cut      & --         &  0.02\%    & $<0.01\%$    \\
Capture Time Cut      & 98.6\%     &  0.12\%    & 0.01\%       \\
Gd Capture Ratio      & 83.8\%     &  0.8\%     & $<0.1\%$     \\
Spill-in              & 105.0\%    &  1.5\%     &  0.02\%      \\
Livetime              & 100.0\%    &  0.002\%   & $<0.01\%$    \\
\hline
Combined              &  78.8\%    &  1.9\%     &  0.2\%       \\
\hline
\hline
\end{tabular}
\end{center}
\end{table}

\section{Backgrounds}
Despite the resolving power of the coincidence IBD selection, a few background effects can mimic an IBD signal.  First, independent single events can occur with small time-separation, giving an accidental coincidence.  These events are statistically subtracted by calculating the expected rate of accidentals using the rate of prompt-like single events, delayed-like single events, and the coincidence interval.

Second, a fast cosmogenic neutron can mimic an IBD signal by first scattering inside the detector to give prompt light and then capturing on Gd to give a delayed signal. 
Since the neutron energy deposition is expected to be flat below 50 MeV, these neutron events can be subtracted from the IBD candidates by considering the prompt-like signal energies with 12-50 MeV and extrapolating into the IBD signal region 0.7-12 MeV.  This method is validated by examining fast-neutron events tagged with a muon.

Third, long-lived cosmogenic isotopes can create an IBD-like coincidence.  For example, ${^9{\rm Li}} \rightarrow {^9{\rm Be}} + e^- + \bar{\nu_e}$ with half-life {178}~{ms}, followed by ${^9{\rm Be}}\rightarrow n+2\alpha$ yields prompt light from the electron and a delayed neutron capture.  This background is studied vs. time after a muon and controlled to $\approx$0.2\% of the IBD signal.

Fourth, the 0.5-Hz ${^{241}{\rm Am}}-{^{13}{\rm C}}$ neutron sources introduce a background when they are retracted into the automated calibration units on the top of the AD.  The neutron scatters off of iron, then captures on stainless steel, producing correlated prompt and delayed signals corresponding to $0.3\pm0.3\%$ of the IBD rate at the far site.

Other backgrounds result from alpha and neutron interactions with carbon and iron in the detector materials.
Although the uncertainties on these contributions are high, these are negligible effects overall.  A summary of the background contributions are given in Table~\ref{tbl:dataset}.

\section{Results}

\begin{center}
\begin{table*}[ht]
\caption{Summary of the Dec. 24 -- May 11 data set.  Where possible, numbers are given for each hall or site.  The upper half of the table describes the raw data in terms of total IBD candidates before background subtraction, data acquisition (DAQ) livetime and efficiency.  The efficiency is dominated by the muon veto rate, which varies with the overburden at each site.  The next five rows give the backgrounds per day, which are subtracted from the IBD candidates to yield the total IBD rate per day in the bottom row.}\label{tbl:dataset}
\begin{tabular}{|l|c|c|c|c|c|c|}
\hline
 & \multicolumn{3}{|c|}{\bf Near Sites} & \multicolumn{3}{|c|}{\bf Far Site} \\ \hline 
 & \multicolumn{2}{|c|}{\bf EH1} & {\bf EH2} & \multicolumn{3}{|c|}{\bf EH3} \\ \hline 
 & ~~AD1~~ & ~~AD2~~ & ~~AD3~~ & ~~AD4~~ & ~~AD5~~ & ~~AD6~~ \\ \hline
IBD Candidates & 69121 & 69714 & 66473 & 9788 & 9669 & 9452 \\ \hline
DAQ livetime (hours) &  \multicolumn{2}{|c|}{127.5470} & 127.3763 &  \multicolumn{3}{|c|}{126.2646} \\ \hline
Efficiency & 0.8015 & 0.7986 & 0.8364 & 0.9555 & 0.9552 & 0.9547 \\ \hline
\hline
Accidentals (/day)       &  9.73$\pm$0.10  & 9.61$\pm$1.10  & 7.55$\pm$0.08  &   3.05$\pm$0.04  & 3.04$\pm$0.04  & 2.93$\pm$0.03  \\ \hline
Fast neutrons (/day)      & \multicolumn{2}{|c|}{0.77$\pm$0.24} & 0.58$\pm$0.33 & \multicolumn{3}{|c|}{0.05$\pm$0.02} \\ \hline
$^{8}$He/$^{9}$Li (/AD/day)       & \multicolumn{2}{|c|}{2.9$\pm$1.5} & 2.0$\pm$1.1 & \multicolumn{3}{|c|}{0.22$\pm$0.12} \\ \hline
AmC (/AD/day)                     & \multicolumn{6}{|c|}{0.2 $\pm$ 0.2 } \\ \hline
$^{13}$C($\alpha$,n)$^{16}$O (/AD/day) & 0.08$\pm$0.04 & 0.07$\pm$0.04 & 0.05$\pm$0.03 & 0.04$\pm$0.02 & 0.04$\pm$0.02 & 0.04$\pm$0.02 \\ \hline
\hline
Antineutrino Rate (/day) & 662.47$\pm$3.00 & 670.87$\pm$3.01 & 613.53$\pm$2.69 & 77.57$\pm$0.85 & 76.62$\pm$0.85 & 74.97$\pm$0.84 \\ \hline
\end{tabular}
\end{table*}
\end{center}

Overall, the Daya Bay experiment collected over 230,000 IBD candidates during the running period from December 24, 2011 to May 11, 2012.  
The antineutrino rates are consistent within each experimental hall, and the dominant uncorrelated detector systematic uncertainty is the delayed energy cut at 6~MeV.

\begin{figure}[ht]
   \centering
   \includegraphics[width=0.95\columnwidth,trim=3 3 3 3,clip]{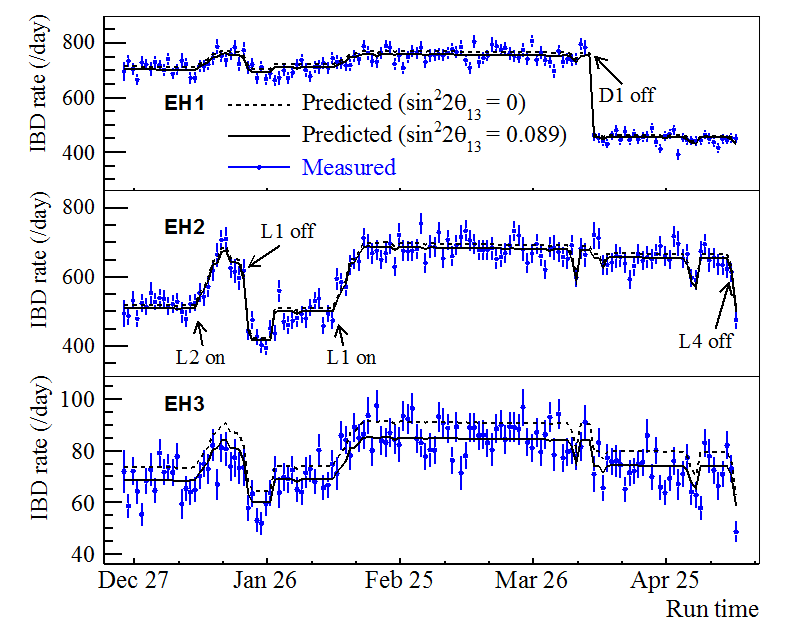} 
   \caption{The measured neutrino events (points) are in good agreement with the prediction from reactor power (lines) for $\sin^2 2\theta_{13}=0.089$.  At the far site (EH3), the null oscillation hypothesis clearly disagrees with the data.
   \label{fig:reactor_vs_time}}
\end{figure}

The event rates at the near sites are weighted by reactor power and baseline to predict the event rate at the far site under a null-oscillation hypothesis.  The ratio of far to near events is 
\begin{equation}
R=0.944 \pm 0.007(\mathrm{stat}) \pm 0.003(\mathrm{syst}).
\end{equation}
The event rate vs. day 
at the three experimental sites tracks the reactor power, and a clear deficit is visible for the far site (Figure~\ref{fig:reactor_vs_time}).

\begin{figure}[ht]
   \centering
   \includegraphics[width=0.95\columnwidth,trim=3 3 3 3,clip]{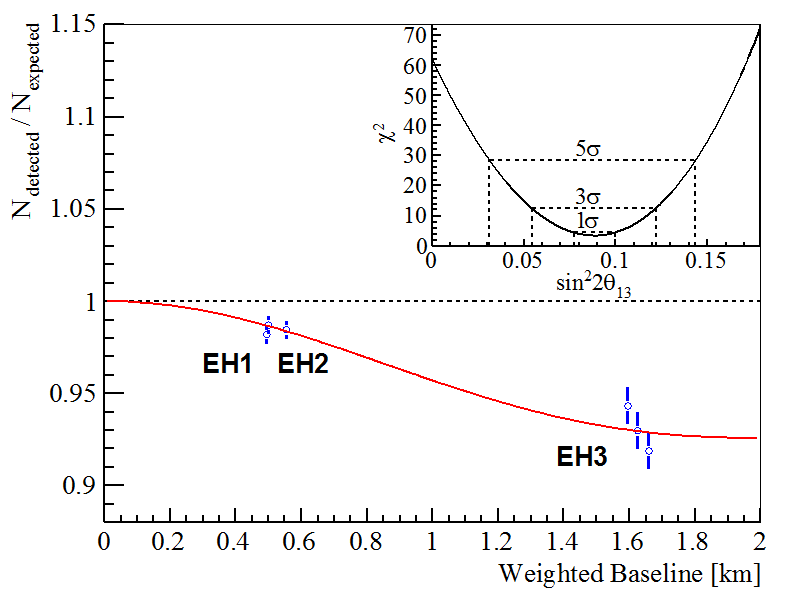} 
   \caption{The ratio of predicted to observed events at the six detectors vs. flux-weighted baseline is fit for $\sin^2 2\theta_{13}$.  The $\chi^2$ plot in the upper right shows $\sin^2 2\theta_{13}=0$ is excluded by over $5\sigma.$
   \label{fig:rate_analysis}
}
\end{figure}

\begin{figure}[ht]
   \centering
   \includegraphics[width=0.95\columnwidth,trim=3 3 3 3,clip]{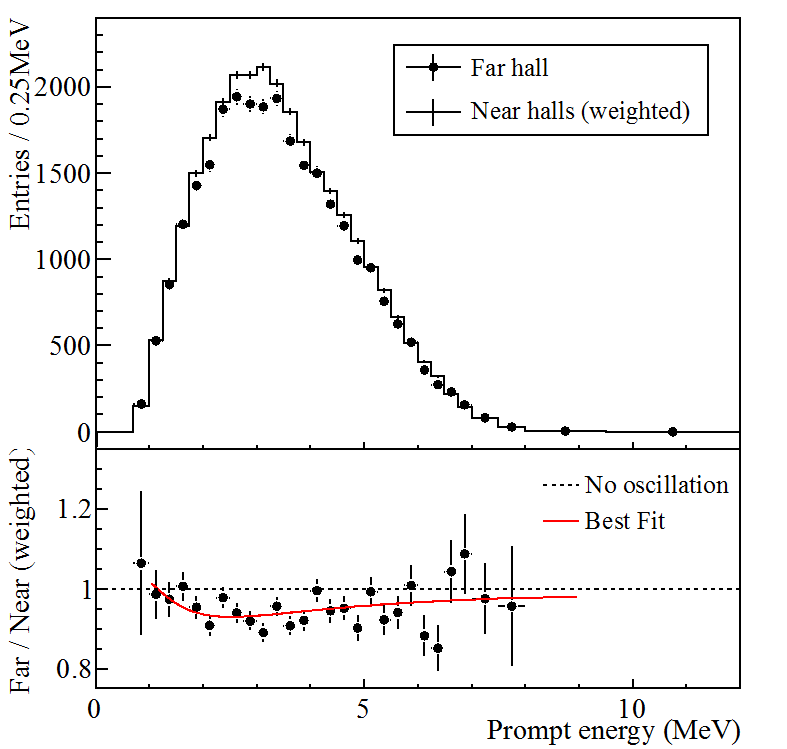} 
   \caption{The upper plot shows the prompt energy spectrum at the near and far experimental halls, weighted by the reactor power and baseline.  The lower plot shows the ratio, with the best-fit rate-only analysis curve superimposed.}
   \label{fig:prompt}
\end{figure}

A relative rate-only analysis was performed on the IBD rates in the 6 detectors, independent of any reactor flux model.  In the standard 3-neutrino framework, a fit for $\theta_{13}$ yields $\sin^2 2\theta_{13} = 0.089 \pm 0.010(\mathrm{stat}) \pm 0.005(\mathrm{syst})$ (Figure~\ref{fig:rate_analysis}).  A superposition of the predicted and observed neutrino events with energy, overlaid with the rate-only best-fit line, is shown in Figure~\ref{fig:prompt}.

\section{Future}
During the summer of 2012, the Daya Bay experiment will install the final two antineutrino detectors. 
 A series of special calibration measurements will be made using radioactive sources deployed along strings from the top of the detector and on an arm inserted into the central volume of one detector.  These measurements will improve the understanding of detector uniformity and energy resolution.  
 
Continuing its science mission over the next 2--3 years, Daya Bay will make a definitive and precise measurement of $\sin^2 2\theta_{13}$.  In addition, the location of the oscillation maximum in the positron energy spectrum will allow a measurement of $\Delta m^2_{ee}$, a combination of $\Delta m^2_{31}$ and $\Delta m^2_{32}$.

In addition, Daya Bay will pursue several secondary scientific and technical studies.  
With the largest antineutrino dataset ever collected, Daya Bay will make the best reactor antineutrino flux and spectra measurements.  Daya Bay will also measure the rates of cosmogenic neutron and isotope production at a range of modest depths and energies.  
In terms of technical studies, Daya Bay will demonstrate the multi-year operation of functionally identical detectors and measure long-term GdLS stability. Finally, the Day Experiment has the potential to search for non-standard neutrino interactions \cite{Dwyer:2011xs}.
Although Daya Bay has produced the most precise value of $\sin^2 2\theta_{13}$ to date, there are many more results still to come.

\section{Acknowledgements}
The Daya Bay experiment is supported in part by
the Ministry of Science and Technology of China,
the United States Department of Energy, the Chinese
Academy of Sciences, the National Natural Science
Foundation of China, the Guangdong provincial government, the Shenzhen municipal government, the China
Guangdong Nuclear Power Group, Shanghai Laboratory for Particle Physics and Cosmology, the Research
Grants Council of the Hong Kong Special Administrative Region of China, University Development Fund of
The University of Hong Kong, the MOE program for
Research of Excellence at National Taiwan University,
National Chiao-Tung University, and NSC fund support
from Taiwan, the U.S. National Science Foundation, the
Alfred P. Sloan Foundation, the Ministry of Education,
Youth and Sports of the Czech Republic, the Czech Science Foundation, and the Joint Institute of
Nuclear Research in Dubna, Russia. We thank Yellow River Engineering Consulting Co., Ltd. and China railway 15th
Bureau Group Co., Ltd. for building the underground
laboratory. We are grateful for the ongoing cooperation
from the China Guangdong Nuclear Power Group and
China Light \& Power Company.




\bibliographystyle{elsarticle-num}
\bibliography{Webber_Beach2012}







\end{document}